# Radial Distributions of Power and Isotopic Concentrations in Candidate Accident Tolerant Fuel $U_3Si_2$ and $UO_2/U_3Si_2$ Fuel Pins with FeCrAl Cladding

Shengli Chen[1,2], Cenxi Yuan[1,*], and Daxi Guo[3]

[1] Sino-French Institute of Nuclear Engineering and Technology, Sun Yat-sen University, Zhuhai, Guangdong 519082, China
[2] CEA, Cadarache, DEN/DER/SPRC/LEPh, 13108 Saint Paul Les Durance, France
[3] China Nuclear Power Technology Research Institute, Shenzhen, Guangdong 518027, China
*Corresponding author. *E-mail address:* yuancx@mail.sysu.edu.cn

## Abstract

Monte Carlo simulations show similarity on radial distributions of power and isotopic concentrations at any effective full power depletion time among five kinds of fuel-cladding combinations with the same cycle length, including the normal $UO_2$-zircaloy combination, the candidate Accident Tolerant Fuel (ATF) $UO_2/U_3Si_2$-FeCrAl combination, and three kinds of candidate ATF $U_3Si_2$-FeCrAl combinations. An analytical formula f($x,s$) including the fuel exposure ($s$) and the relative radial ($x$) is proposed to describe the radial properties for all five kinds of fuel-cladding combinations. f($x,s$) has the form of the second order polynomial term of $s$ with the exponential type of coefficients depending on $x$. It is shown that the suggested function f($x,s$) gives a nice description on the simulation data with rather small deviations and can immediately provide radial distribution of power, burnup, and isotopic concentrations of $^{235}U$, $^{238}U$, $^{239}Pu$, and $^{241}Pu$ at any fuel exposure and relative radius. It is useful to discuss the fuel temperature through the present analytical formula. The realistic radial power distribution gives flatter radial temperature distribution compared with the uniform power distribution. Because of the different thermal conductivities of fuels and claddings and the different thicknesses of claddings, the present discussed five kinds of fuel-cladding combinations have different radial temperature distributions, although their radial power distributions are quite similar. The present work provides an analytical formula to describe the radial properties of the ATF which is expected to be helpful for further neutronic and multi-physics coupling studies.

## 1. Introduction

Because of the Fukushima nuclear disaster in 2011, the Accident Tolerant Fuel (ATF) has been developed for advanced nuclear fuel and cladding options. The US DOE NE Advanced Fuels Campaign has proposed many candidate ATF fuels and claddings [1]. As explained in Ref. [2], FeCrAl has better accident tolerant performance than current zircaloy cladding. The welding of FeCrAl is investigated with both the fusion-based laser and the solid-state joining with pressure resistance [3]. In order to compensate its larger neutron capture cross section, the $U_3Si_2$ fuel has been proposed because of its larger density and higher uranium concentration. The $U_3Si_2$ fuel is also the candidate ATF suggested by Westinghouse [4], [5]. Westinghouse has studied the stability under coolant conditions [6].

The neutronic studies showed that the $U_3Si_2$-FeCrAl can be a potential ATF combination [2], [7].

Besides $U_3Si_2$ fuel, the mixtures of uranium-mononitride (UN) and uranium silicides, such as $U_3Si_2$(32%)/UN(68%) [8] and $U_3Si_5$/UN [9], are also candidate ATFs. However, about 90% enrichment of $^{15}N$ in UN is needed in application due to the large thermal neutron capture cross section of $^{14}N$ [10]. In addition, UN chemically reacts with water at high temperature [11]. Westinghouse also proposes $UO_2$/$U_3Si_2$ mixed fuel combination [12]. $UO_2$/$U_3Si_2$-FeCrAl may be a candidate fuel-cladding combination because both $UO_2$/$U_3Si_2$ fuel and FeCrAl cladding are proposed. According to the design of semi-heterogenous loading for minors actinides transmutation [13], [14], it is possible to propose an assembly including both the $UO_2$ and $U_3Si_2$ fuel rods. The present work studies the relationship between the reactivity at the End of Cycle (EOC) and the percentage of $UO_2$ in the homogeneously mixed $UO_2$/$U_3Si_2$ fuel.

On the other hand, the periphery phenomena in fuel rod are evident due to the spatial self-shielding in Pressurized Power Reactors (PWR) [15]. Radial properties of a fuel pellet have been investigated. For example, Lassmann and coworkers have studied the radial distribution of plutonium in the $UO_2$ pellet in 1994 [16]. In the last five years, extensive focuses on radial distributions have been developed. For example, a phenomenological study is performed on the radial power density, the isotopic concentration, the burnup, and other properties of the $UO_2$ fuel [17]. The radial structure of a fuel pellet at high burnup level is also analyzed [18]. Pirouzmand and Roosta investigated the radial distribution of the burnup and the atomic density in the VVER-1000 fuel [19]. We have also found a formula in simple form for the radial distributions of burnup and reaction rate in the $UO_2$ fuel [20]. It is important to investigate the radial distribution of physical properties for a new fuel-cladding combination. The present work focuses on the radial distributions of power and isotopic concentration of the $UO_2$-zircaloy, $U_3Si_2$-FeCrAl, and $UO_2$/$U_3Si_2$-FeCrAl systems.

The radial properties of the present considered ATFs are affected by many fuel variables, including the fuel exposure ($s$), the relative radius ($x$), the uranium enrichment ($e$), the thickness of the FeCrAl cladding ($t$), and the volume fraction of $UO_2$ in the $UO_2$/$U_3Si_2$ fuel ($v$). $t$ is normally determined by the cladding material. Because $t$ and $e$ have a synergic effect on the reactivity, $e$ can be determined for a given reactor to meet the requirements of power and cycle length. For the $UO_2$/$U_3Si_2$ fuel case, one more factor $v$ needs to be fixed.. The determination of $v$ for a given system is shown in section 3. The last three parameters can be well determined for each system. The present work focuses on the effect of the first two variables on the radial distributions of power and isotopic concentration because of the impossibility of including all variables and the continuous variation of $s$ and $x$. In addition, five cases studied in this work show the similar radial neutronic properties, while $t$, $e$, and $v$ are different (shown in sections 2 and 3).

At the beginning of section 2, we give a brief introduction to the simulation methods. An analytical formula is proposed to describe the radial distributions of power and isotopic concentration at any radial position and burnup level. Section 3 shows the simulation results and the fitting and verification of the suggested analytical formula. The radial distribution of the fuel temperature is discussed based on the formula at the end of this section. The discussions and conclusions about the analytical expression of the radial distributions of power and isotopic concentration are presented in section 4.

## 2. Method

2.1 Simulation methods

The Monte Carlo methods are normally more reliable than the deterministic methods for a new fuel-cladding combination. For example, the existing formulae in the deterministic methods may provide unsuitable self-shielding corrections for a new system. In the present work, the Monte Carlo based code RMC is used [21]. RMC is a 3D Monte Carlo neutron transport code developed by Tsinghua University. Similar to most Monte Carlo code, RMC is able to deal with complex geometry. The continuous-energy pointwise cross sections for different materials are used in RMC. It has both criticality and burnup calculations, which provide both the effective multiplication factor and the isotopic concentrations at different burnup levels. The simulations performed in the present study take 3.33 Effective Full Power Days (EFPDs), 13.33 EFPDs, and 16.67 EFPDs for the first three steps and 33.33 EFPDs from the fourth step in burnup calculations. Each step contains 10 sub-steps for depletion calculations. All calculates are based on the ENDF/B-VII.0 [22] nuclear data library.

Table 1 Model specification [2]

| Property | Unit | Value |
|---|---|---|
| Assembly fuel height | cm | 365.76 |
| Cladding composition | wt% | Zr-4: Fe/Cr/Zr/Sn = 0.15/0.1/98.26/1.49 * |
| | | FeCrAl: Fe/Cr/Al = 75/20/5 |
| Gap thickness | μm | 82.5 |
| Cladding thickness | μm | Zircaloy-4: 571.5 * |
| Cladding Outer Radius | mm | 4.750 |
| Fuel enrichment | % | 4.9 |
| Pitch to outer Diameter | - | 1.326 |
| Cladding IR of guide tube | mm | 5.624 |
| Cladding OR of guide tube | mm | 6.032 |
| Number of guide tubes | - | 25 |
| Fuel density | g/cm$^3$ | $U_3Si_2$: 11.57 |
| | | $UO_2$: 10.47 |
| Specific power density ** | MW/MtU | 38.33 * |
| Coolant density | g/cm$^3$ | 0.7119 |
| Helium density | g/L | 1.625 (2.0 MPa) |
| Cladding density | g/cm$^3$ | Zircaloy-4: 6.56 |
| | | FeCrAl: 7.10 |
| Coolant temperature | K | 580 |
| Fuel temperature | K | 900 |

| Cladding and gap temperature | K | 600 |
|---|---|---|
| Boron concentration | ppm | 630 |
| Boundary conditions | - | Reflective |

* Values only for the reference case
** Specific power density defines the nuclear power of unit mass uranium. It is calculated by keeping the total power in a fuel assembly, so that the power in the core does not change when different designs of fuel assembly are used.

Table 2 Distribution of population and power per fuel cycle batch in a typical Westinghouse PWR [23]

| Batch | Number of assemblies | Core fraction vol% ($V_b$) | Relative assembly power ($P_b$) | EFPDs achieved at EOC ($e_b$) |
|---|---|---|---|---|
| 1 | 73 | 38% | 1.25 | 627 |
| 2 | 68 | 35% | 1.19 | 1221 |
| 3 | 52 | 27% | 0.40 | 1420 |
| Total | 193 | 100% | - | - |

The 4.9% uranium enrichment for both UO$_2$ and U$_3$Si$_2$ are the same as a typical Westinghouse PWR, of which all parameters are listed in Table 1. The FeCrAl cladding is used for UO$_2$/U$_3$Si$_2$ mixed fuel study. A 350μm thickness FeCrAl cladding is used for the UO$_2$/U$_3$Si$_2$ fuel because the same thickness stainless steel was used in Light Water Reactor (LWR) [24]. The criterion to choose the percentage of UO$_2$ is that the UO$_2$/U$_3$Si$_2$-FeCrAl combination ensures the same cycle length as the normal UO$_2$-Zircaloy combination in a typical Westinghouse PWR. The formula to calculate the average difference of reactivity in a reactor core at the EOC is [25]:

$$\Delta k_{core} = \frac{\sum_b \Delta k_{inf,b}(e_b) P_b V_b}{\sum_b P_b V_b}, \qquad (1)$$

where $\Delta k_{inf,b}$ is the difference of infinity multiplication factor $k_{inf}$ between the fuel design under consideration and the reference case for batch $b$ of the fuel assembly as a function of exposure $e_b$. The calculations of $k_{inf}$ are performed on a fuel assembly with parameters given in Table 1. The EOC fuel exposure (in EFPDs) from Table 2 are used to quantify the level of exposure for each batch. The power weighting factor $P_b$ approximates the power distribution in the core to provide the contribution of each batch. The number of assemblies per batch is denoted as $V_b$.

Once the percentage of UO$_2$ in UO$_2$/U$_3$Si$_2$ mixed fuel is obtained, the present work investigates the radial power distribution for current UO$_2$-zircaloy combination (noted as Ref in following), obtained UO$_2$/U$_3$Si$_2$-FeCrAl combination, and the U$_3$Si$_2$-FeCrAl combinations with 571.5μm, 450μm, and 350μm cladding thicknesses. The 571.5μm thickness corresponds to the current zircaloy cladding thickness. The 450μm thicknesses of FeCrAl is the thickness keeping the same cycle length as the reference case when the 4.9% uranium enrichment U$_3$Si$_2$ fuel is used [2]. The 350μm thicknesses of stainless steel was used in LWR [24]. The fuel enrichment of each combination is determined through the Eq. (3) in Ref. [2] and given in Table 3.

Table 3 Main parameters

| Case | Fuel | Fuel density | Uranium enrichment | Cladding | Cladding thickness |
|---|---|---|---|---|---|
| Ref | $UO_2$ | 10.47g/cm$^3$ | 4.90% | Zircaloy | 571.5μm |
| t=571.5μm | $U_3Si_2$ | 11.57g/cm$^3$ | 5.29% | FeCrAl | 571.5μm |
| t= 450μm | $U_3Si_2$ | 11.57g/cm$^3$ | 4.90% | FeCrAl | 450μm |
| t=350μm | $U_3Si_2$ | 11.57g/cm$^3$ | 4.58% | FeCrAl | 350μm |
| $UO_2/U_3Si_2$ | $UO_2/U_3Si_2$ | g(v) g/cm$^3$* | 4.90% | FeCrAl | 350μm |

* $g(v) = v\rho(UO_2) + (1-v)\rho(U_3Si_2)$, where $v$ and $\rho$ represent volume fraction of $UO_2$ in mixed fuel and masse density respectively.

The geometry of the present fuel rods is the same as the proposed one in Ref. [2]. As shown in Figure 1, the fuel region is divided into 9 rings, while the gap and cladding are located outside. The distances between two rings are chosen to be denser when they are closed to the periphery of the fuel rods. The power distribution is rather flat in the middle of the fuel rod, but increases very quickly when approaching to the periphery of the fuel rod. In addition, the small rings need much more computational cost to reduce the uncertainty of the simulation. In the present study, the thickness of the most outside ring is only $0.014r_0$ (around 0.006 cm), which is much smaller than the mean free paths of all neutrons. Only few neutrons react in such small volume, which increases the uncertainty of the Monte Carlo simulation a lot. The present set, of which the rings are sparse in the middle and dense in the periphery, is chosen to balance the computational cost and the accuracy.

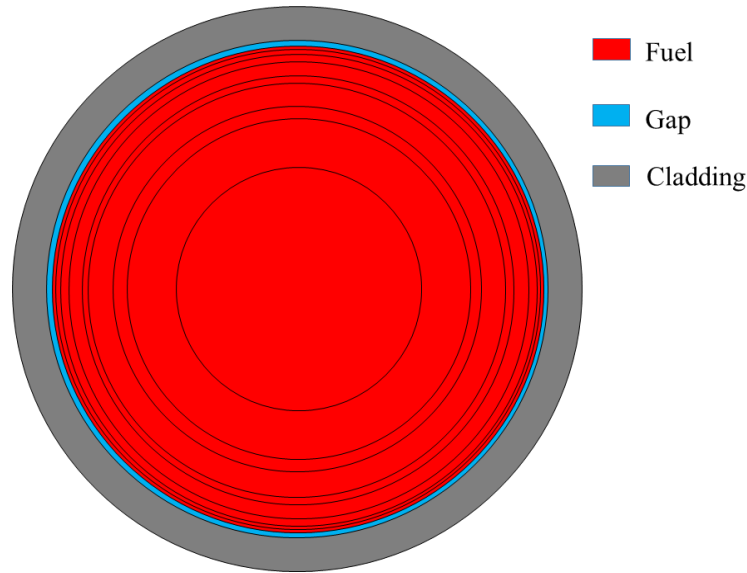

Figure 1. Radial profile of a fuel-cladding system with fuel region divided into 9 rings

Figure 2 points out the radial power distributions computed by a Finer mesh (13 rings noted by F) and the Normal mesh (shown in Figure 1, noted by N) for the reference case at different fuel exposures. The agreement between two results shows that the geometry shown in Figure 1 is appropriate to compute the radial distributions of power and isotopic concentrations. In addition, the reasonable radial distributions shown in Section 3 and the close to unity values of renormalization factor of power distribution (given in Section 3.5) further confirm that the proposed mesh is suitable to compute the radial distributions of power and isotopic concentrations.

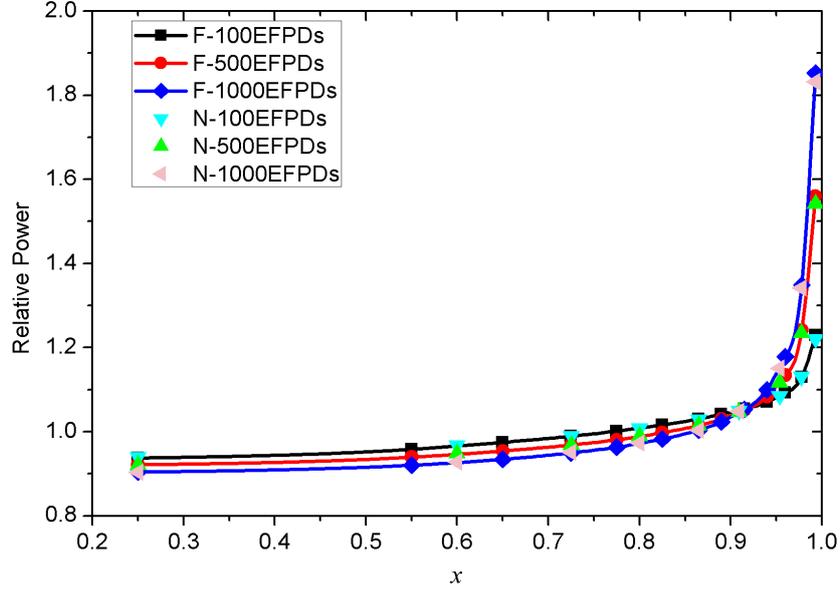

Figure 2. Radial power distributions computed by a Finer mesh (13 rings noted by F) and the Normal mesh (shown in Figure 1, noted by N) for the reference case at different fuel exposures

2.2 Radial distributions

Lassmann and coworkers proposed a general form for radial properties distributions at radius $r$ as [16]:

$$f(r) = p_0 + p_1 \exp(-p_2(r_0 - r)^{p_3}), \quad (2)$$

where $r_0$ is the radial of fuel pellet, $p_i (i = 0,1,2,3)$ are constants obtained by fitting. This formula has been used in the burnup calculation [26]. From the expression, it is evident that the parameters $p_2$ and $p_3$ are very important to describe the radial distribution and need high accuracy. It is not easy to obtain high accurate parameters through the optimization of the exponential term in Eq. (2), especially for $p_3$. On the other hand, $p_i (i = 0,1,2,3)$ depends on the burnup level. It is difficult to obtain a huge amount of high accurate parameters to reproduce the radial power distribution.

In order to overcome the shortcoming of fitting at different burnup levels of Lassmann's formula, the following method is proposed to reproduce the radial power distribution. In fact, the radial power distribution as a function of the fuel exposure at a given spatial position can be well described by a second order polynomial:

$$f(x, s) = a(x)s^2 + b(x)s + c(x), \quad (3)$$

where $s$ represents the fuel exposure in EFPDs. The relationship between $s$ and burnup ($bu$) is $bu = s \cdot SPD$, where $SPD$ is the Specific Power Density (in MW/MtU). The $SPD$ for the reference case is shown in Table 1 and can be computed for other cases by keeping the total power in a fuel assembly. $s$ is used in the present work because the similar radial power distributions are observed for different cases (see section 3). The coefficients of polynomial depend on the relative radius $x$, where $x = r/r_0$ and $r_0$ represents the radius of fuel pellet. Eq. (3) is also used to describe the relative isotopic concentration for important actinides, such as $^{235}$U, $^{238}$U, $^{239}$Pu, and $^{241}$Pu.

For each radial point, only 3 coefficients are needed to determine the relative radial power distribution. When relative power distribution is obtained, the local burnup can be calculated by:

$$\frac{bu(x,s)}{\overline{bu}} = \frac{1}{s}\int_0^s f(x, s')ds', \qquad (4)$$

where $\overline{bu}$ is the average burnup, which is proportional to $s$. Inserting Eq. (3) to Eq. (4):

$$\frac{bu(x,s)}{\overline{bu}} = \frac{1}{3}a(x)s^2 + \frac{1}{2}b(x)s + c(x). \qquad (5)$$

The interpolation method can be used to determine the radial distributions of power and isotopic concentration at any burnup level when the coefficients in Eq. (3) are obtained. In fact, these coefficients can be further expressed as the following form:

$$i(x) = A\exp(B(1-x)) + C\exp(D(1-x)) + E, \qquad (6)$$

where $i(x) = a(x), b(x), c(x)$, and $A, B, C, D, E$ are constants which will be determined by radial distributions of $a, b, c$. A similar form is also used in the description of the radial reaction rate in Ref. [20].

More accurate temperature distribution in the fuel pellet can be calculated through the radial power distribution. In a fuel rod, when neglecting the thermal radiation, the differential Equation of the fuel temperature $T$ in steady state is:

$$\frac{1}{r}\frac{d}{dr}\left(\lambda(T)r\frac{dT}{dr}\right) + f(r)Q = 0, \qquad (7)$$

where $\lambda$ is the thermal conductibility, in W/K/m, and $Q$ is the average volumic thermal power. The thermal conductibility depends on temperature. The Fink's empirical expression for UO$_2$ is [27]:

$$\lambda(t) = 100(7.5408 + 17.692t + 3.6142t^2)^{-1} + 6400t^{-2.5}\exp(-16.35/t), \qquad (8)$$

where $t = T/1000$, and that for the U$_3$Si$_2$ fuel is given by [28]:

$$\lambda(T) = 7.98 + 0.0051 \times (T - 273.15). \qquad (9)$$

Eq. (7) shows that the radial power distribution investigated in the present work can be directly used for fuel temperature calculation, which is of high importance for both neutronic and multi-physics coupling calculation. The same total power in a fuel pin implies that $\pi r_0^2 Q$ is the same for all cases. By using the symmetry of $T$ at $r = 0$ and the substitution $x = r/r_0$, double integration of Eq. (7) gives directly:

$$\int_{T(x=1)}^{T(x)} \lambda(T)dT \propto -\int_1^x \left[\int_0^{x'} x''f(x'')dx''\right]/x'dx'. \qquad (10)$$

For radial power distribution at the right side, the analytical Eq. (3) is used. Then the "pchip" interpolation method in Matlab [29] is used to generate full domain distribution from 9 considered points in the present work and $x = 0$, in which the value at $x = 0.25$ is used to avoid unrealistic change near to the center of the fuel rod due to the numerical extrapolation. According to the definition of the normalization factor for relative power distribution, there is

$$\int_0^1 2\pi x f(x)dx = \pi. \qquad (11)$$

By consequence, the factor f(x,s) should be renormalized after the interpolation.

## 3. Results and Discussion

3.1 Multiplication factor

Figure 3 shows the infinity multiplication factors $k_{inf}$ for the reference case and some volume fractions (v) of UO$_2$ in the mixed fuel. Results in Figure 3 is in good agreement with one of the conclusions in Ref. [2]: $k_{inf}$ in the U$_3$Si$_2$ fuel changes less versus depletion time than that in UO$_2$ fuel.

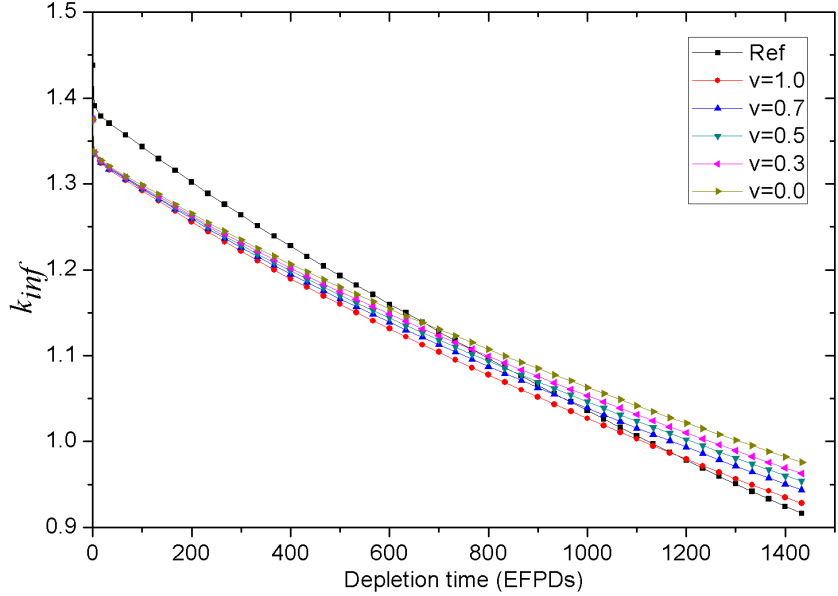

Figure 3. Infinity multiplication factor $k_{inf}$ versus effective full power depletion time in a fuel assembly

As given in Table 4, the $\Delta k_{core}$ values between the UO$_2$(v)/U$_3$Si$_2$(1-v)-FeCrAl combinations and the reference case are calculated for different volume fractions $v$ through Eq. (1) by using the $k_{inf}$ factors. Based on the results in Table 4, a second order polynomial fitting is proposed to describe $\Delta k_{core}$ as a function of $v$:

$$\Delta k_{core} = \alpha v^2 + \beta v + \gamma, \qquad (12)$$

where $\alpha = -0.00566 \pm 0.00027$, $\beta = -0.02866 \pm 0.00029$, $\gamma = 0.02288 \pm 0.00007$, and the coefficient of determination $R^2 = 0.99996$. The uncertainties of fitted parameters in the present work are from the uncertainties of statistical fitting. The excellent fitting results confirm the proposition of Eq. (12). It shows that 70% volume fraction of UO$_2$ in the mixed fuel has the same cycle length as the current fuel-cladding combination. Further discussion is performed on the UO$_2$(70%)/U$_3$Si$_2$(30%) mixture with 4.9% enrichment of uranium and 350μm thickness of FeCrAl cladding. The density of UO$_2$(70%)/U$_3$Si$_2$(30%) mixed fuel is 10.80 g/cm$^3$. In the following discussions, the UO$_2$(70%)/U$_3$Si$_2$(30%)-FeCrAl fuel-cladding combination is referred to UO$_2$/U$_3$Si$_2$-FeCrAl.

Table 4 $\Delta k_{core}$ between UO$_2$/U$_3$Si$_2$-FeCrAl and the reference case

| $v$ | 0.0 | 0.2 | 0.3 | 0.5 | 0.6 | 0.7 | 0.8 | 0.9 | 1.0 |
|---|---|---|---|---|---|---|---|---|---|
| $\Delta k_{core}$ | 0.0229 | 0.0169 | 0.0137 | 0.0072 | 0.0036 | 0.0000 | -0.0036 | -0.0076 | -0.0114 |

3.2 Radial distributions of power

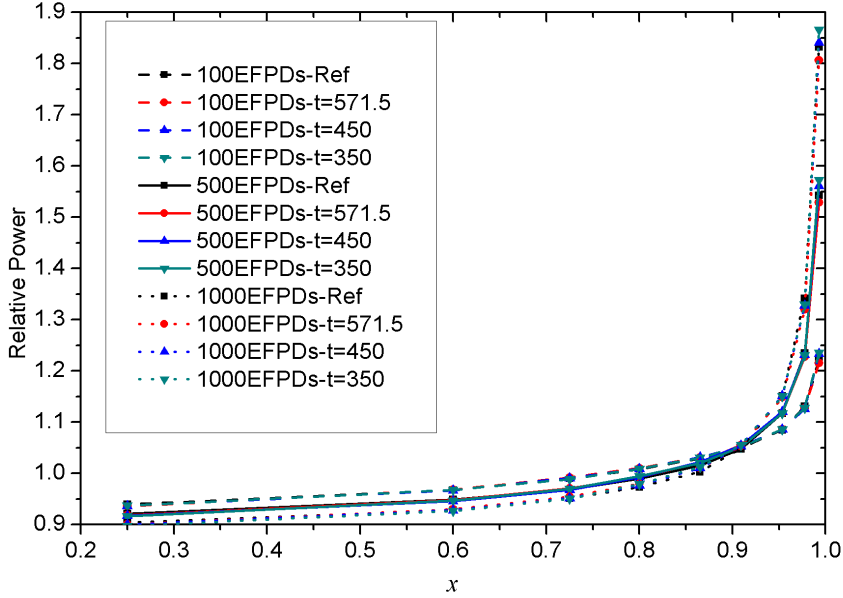

Figure 4. Radial distributions of relative power at 100 EFPDs (dashed lines), 500 EFPDs (solid lines), and 1000 EFPDs (dotted lines) fuel exposure

The radial power distributions of the 5 fuel-cladding combinations are almost the same according to the Monte Carlo simulations. The results at 100 EFPDs, 500 EFPDs, and 1000 EFPDs are shown in Figure 4, where Ref represents the normal UO$_2$-zircaloy combination. The notations, t=571.5μm, 450μm, and 350μm, represent the U$_3$Si$_2$-FeCrAl combination with the corresponding cladding thickness. The notation UO$_2$/U$_3$Si$_2$ represents the above chosen UO$_2$(70%)/U$_3$Si$_2$(30%)-FeCrAl combination. The same notations are used in the following discussions. The relative power near the outer surface of the fuel increases with the fuel exposure because of the fission reaction from $^{239}$Pu. The local power near the periphery is from the (n,γ) reaction of $^{238}$U, which produces $^{239}$Pu and corresponding fission reactions near the periphery. The concentration of $^{239}$Pu increases with the fuel exposure, which will be discussed in the following sub-section of radial distributions of isotopic concentration. Because of the similarity among different cases, a general analytical formula $f(x,s)$ described in section 2 is obtained from the average relative power of the 5 cases. The applicability of the analytical approximation is presented in the following discussion.

The fitting results of $f(x,s)$ are shown in Figure 5, in which the red lines and the points represent the polynomial fitting results and averaged values from the Monte Carlo simulations, respectively. It should be noted that the red curves represent the fitting results for Figure 5 and Figure 10. Table 5 in appendix gives the parameters in Eq. (3), the corresponding relative uncertainties, and the coefficients of determination obtained in the fitting procedure. These results ensure the applicability of the analytical approximation of Eq. (3). The End of Life (EOL) is assumed to be 1420 EFPDs, while the present formula is available up to 2000 EFPDs.

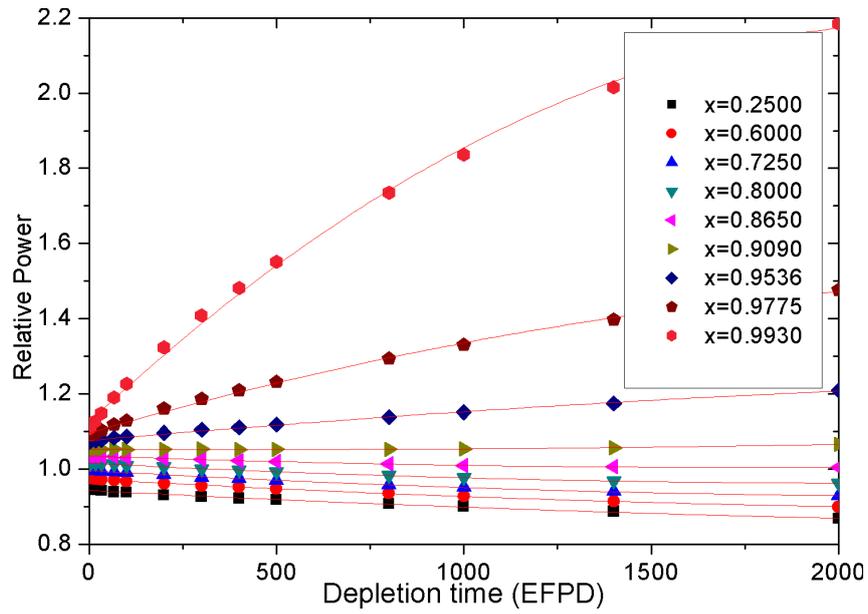

Figure 5. Relative power as a function of depletion time at different positions. The red curves represent the fitting results, the scattering points are simulation data.

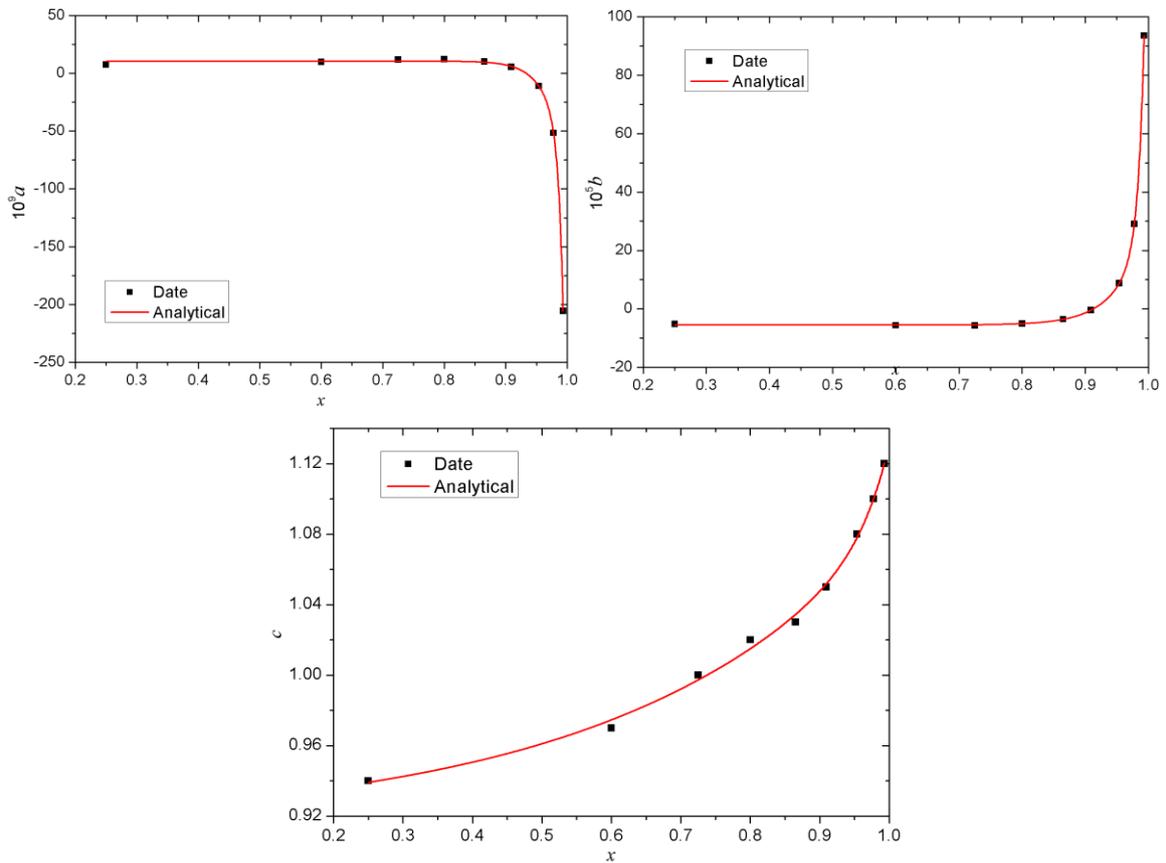

Figure 6. $a(x), b(x),$ and $c(x)$ in Table 5 and the analytical fitting formula. The curves represent the fitting results, the scattering points are data obtained in Table 5.

Constants in Eq. (6), which describe $a(x), b(x),$ and $c(x)$ as a function of the relative radius, are listed in Table 6 in appendix. The coefficients in Eq. (3) can be obtained at any position through Eq. (6) and constants in Table 6. The results are shown in Figure 6.

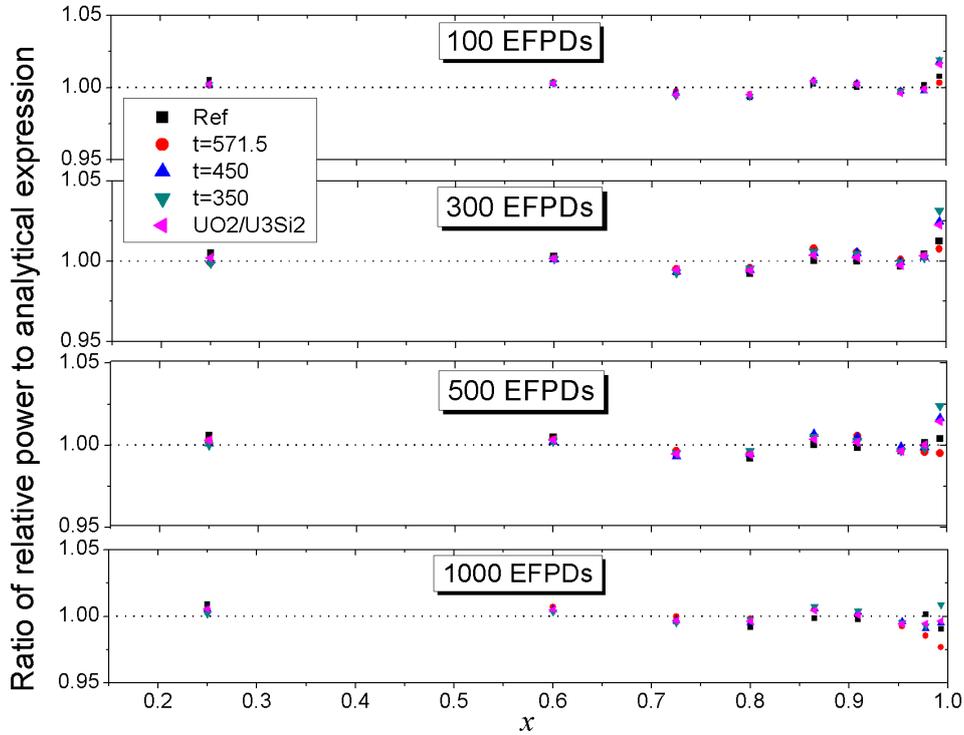

Figure 7. Ratios of power calculated with RMC to that obtained by Eq. (3) at 100 EFPDs, 300 EFPDs, 500 EFPDs, and 1000 EFPDs

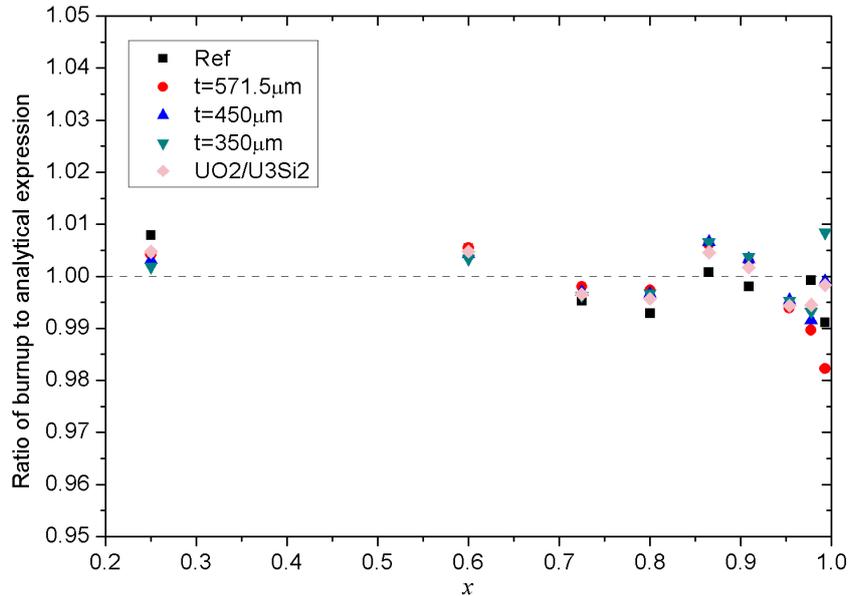

Figure 8. Ratios of burnup calculated with RMC to that obtained by Eq. (5) at the EOL

Once the coefficients are determined, one can calculate relative radial power distribution at any depletion time. Figure 7 shows the ratios of Monte Carlo simulated power to the results obtained by Eq. (3) with corresponding coefficients at 100 EFPDs, 300 EFPDs, 500 EFPDs, and 1000 EFPDs. The fitted equation has almost the largest deviation from the simulated data at 300 EFPDs. The applicability of the present analytical approaches is verified on both degrees of freedom, burnup and radial positions. Figure 8 illustrates the ratios of burnup computed by RMC to Eq. (5) at the EOL. The deviations between Eq. (5) and Monte Carlo simulations are within 2% at the EOL.

It is not necessary to divide the most outside ring to be several smaller rings in the simulation, although the radial power keeps increasing when approaching to the outer surface of the fuel rod. The local burnup phenomena is mainly from the (n,γ) reaction of $^{238}$U (and the corresponding product $^{239}$Pu), of which the cross section is generally very small but can reach to $10^4$ barn ($10^{-20}$ cm$^2$) at certain energies in the resonance region. The concentration of $^{238}$U is of the order of magnitude of $10^{22}$ cm$^{-3}$. The mean free paths of neutrons at such energies are at the magnitude of $1/(10^{-20}*10^{22}) = 0.01$ cm, which is much smaller than the mean free paths of all neutrons. At present, the thickness of the most outside ring is $0.014r_0$ (around 0.006 cm), which reaches the physical limitation of the reaction and no need to be further reduced. In addition, although there are few local burnup measurements for ATF fuel, there are certain data for traditional fuel, such as some experimental data shown in Refs. [16], [18], The minimum radial distance between two neighbor data is also around 0.01 cm. Moreover, because the outer ring is already very small, its total power is also small and has almost no influence on fuel behaviors, such as the fuel temperature distribution discussed in the present work.

3.3 Radial power distribution in perturbated cases

The previous analyses point out the agreement between the Monte Carlo simulations and the empirical formula. As pointed out in section 2, the boron concentration and moderator temperature are constant in our simulations. In real cases, the boron concentration decreases with burnup and the moderator temperature depends on axial position. To check whether the previously observed conclusions can be applied when the moderator temperature and boron concentration change, four perturbated scenarios are computed. The 63 ppm boron concentration and 570 K moderator temperature are separately used in the reference system and the 450 μm thicknesses FeCrAl cladding case. The thermal expansion coefficient is 0.00329871/K [30] for the moderator.

The Monte Carlo simulations are performed for the above four perturbated cases, including the 63 ppm boron concentration and 570 K moderator temperature for two fuel-cladding systems. Figure 9 shows the ratios of Monte Carlo simulated power to the empirical formula obtained with the above fives cases at 100 EFPDs, 500 EFPDs, and 1000 EFPDs burnup level. The results ensure the applicability of the empirical formula to some perturbated scenarios.

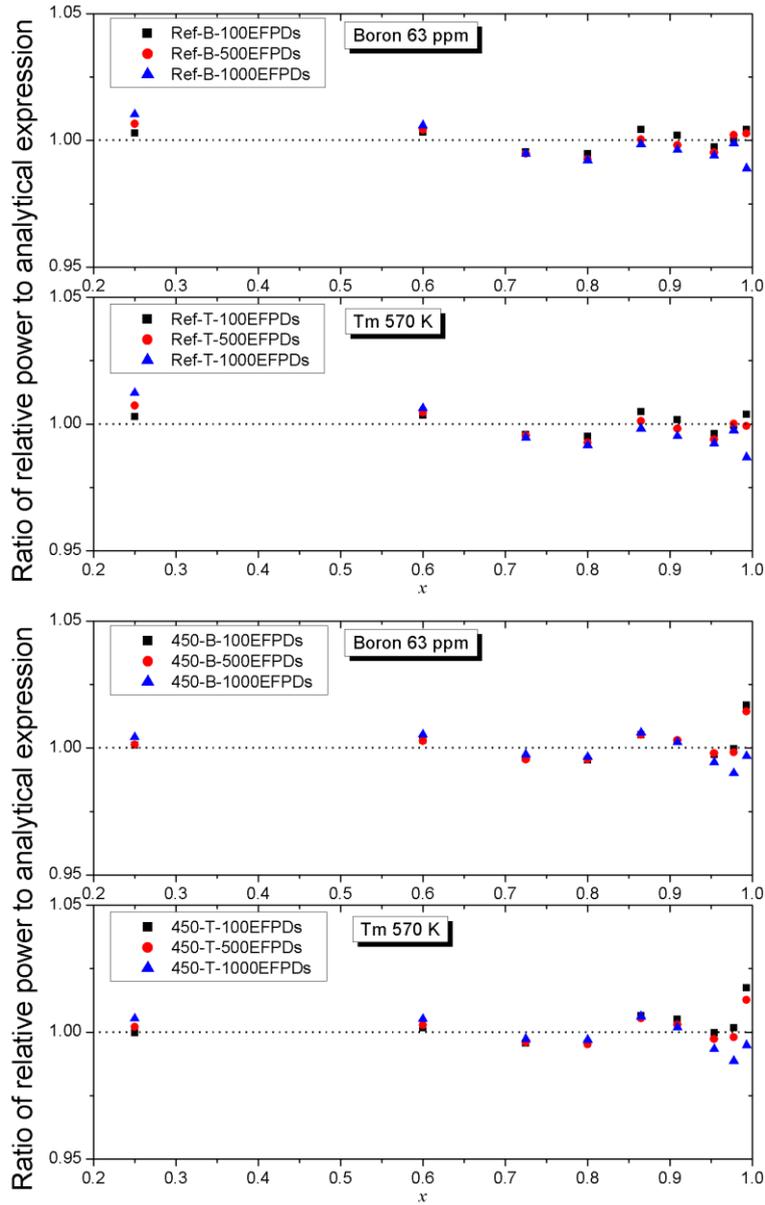

Figure 9. Ratios of power calculated with RMC to Eq. (3) with coefficients in Table 5 at 100 EFPDs, 500 EFPDs, and 1000 EFPDs for the perturbated cases. The first (second) figure are results for the reference (450 μm FeCrAl cladding) case. The up (down) sub-plot for each figure represents the results of the 63 ppm boron concentration (570 K moderator temperature) case.

## 3.4 Radial distributions of isotopic concentrations

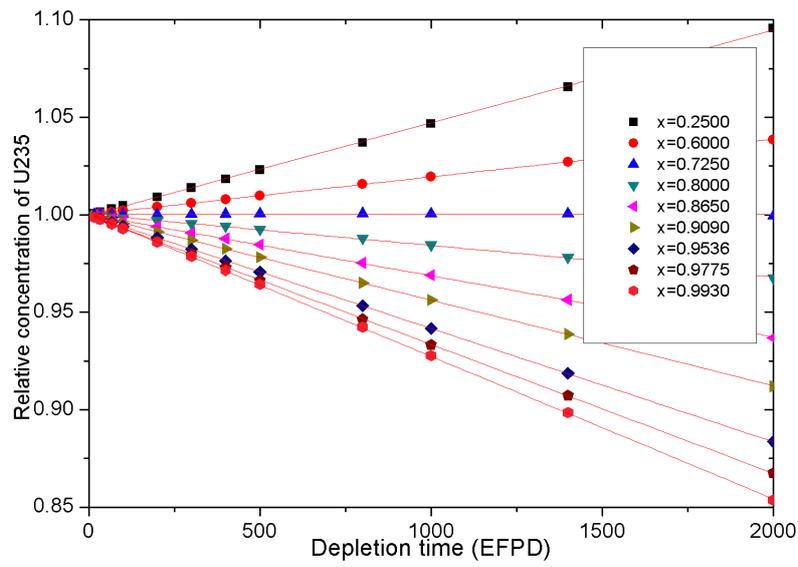

(a)

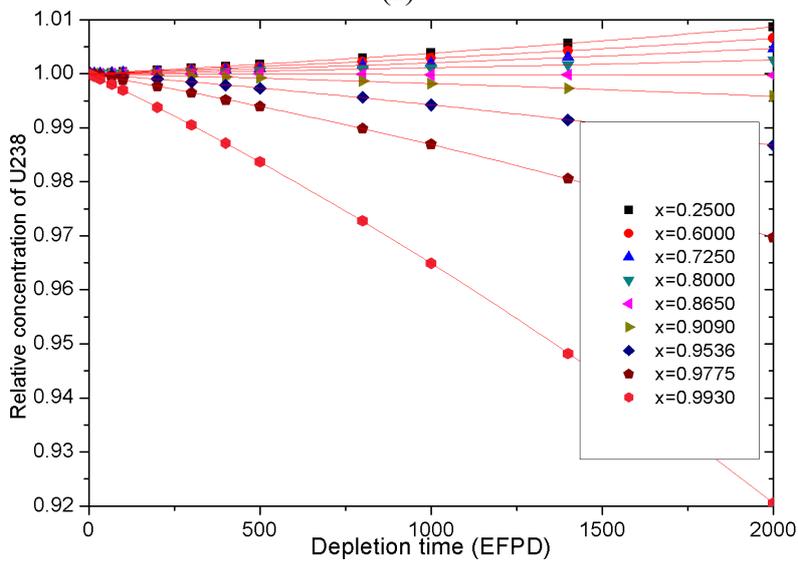

(b)

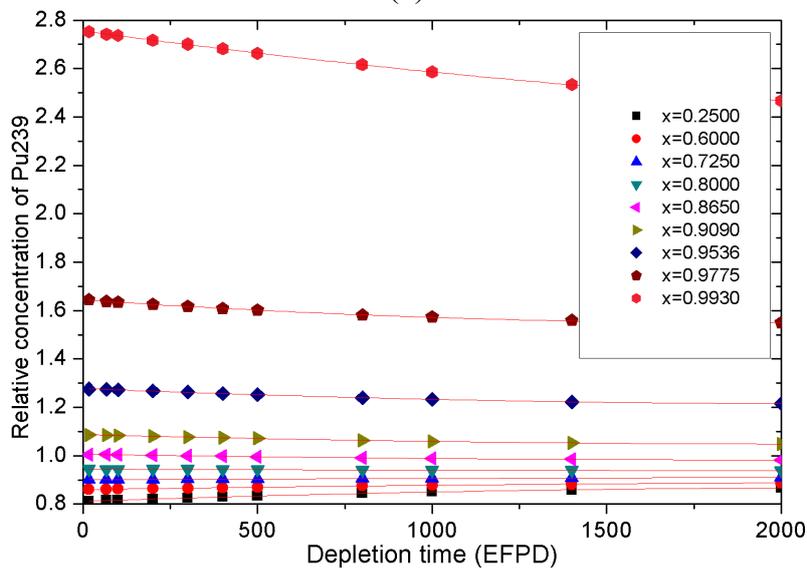

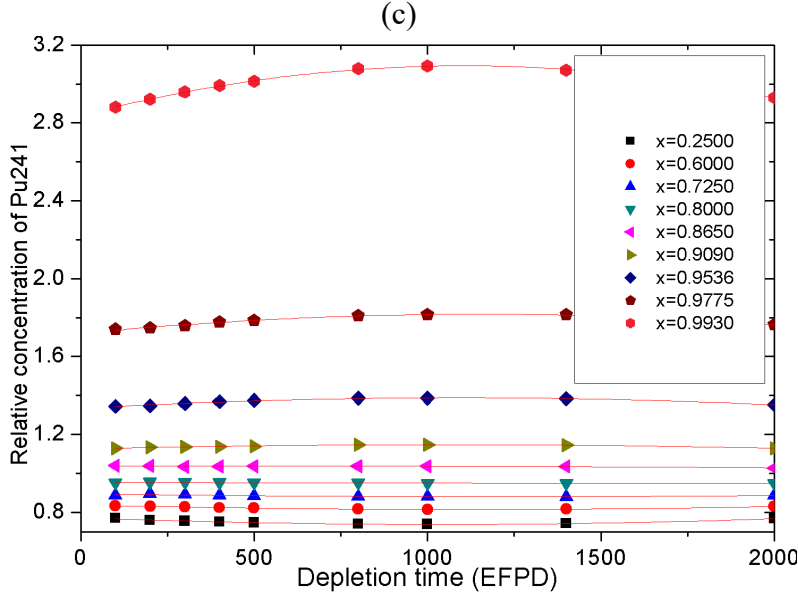

(c)

(d)

Figure 10. Relative concentrations of $^{235}U$, $^{238}U$, $^{239}Pu$, and $^{241}Pu$ for t=450μm case. The red curves represent the fitting results, the scattering points are simulation data.

The relative isotopic concentration is defined as the ratio of local concentration to the average concentration for each isotope at each fuel exposure level. In other words, the relative isotopic concentration of isotope $i$ is defined by $C_i(x,s)/\bar{C_i}(s)$, where $\bar{C_i}(s)$ is the average concentration of isotope $i$ at fuel exposure $s$. The present analytical formula f($x,s$) also gives a nice description on the distribution of relative isotopic concentrations, which are shown in Figure 10 for $^{235}U$, $^{238}U$, $^{239}Pu$, and $^{241}Pu$ in t=450μm case. The corresponding coefficients and associated uncertainties of second polynomials are given in Table 7 in the appendix. In figure 9, the concentrations of $^{235}U$ and $^{238}U$ start from 1.0 because they both have original concentration at the beginning. There are no $^{239}Pu$ nor $^{241}Pu$ at the beginning. They actually start at a small fuel exposure in the simulation. Because of the local power phenomena, their distribution is not uniform along radius at the beginning. Because the burnup (or effective full power depletion time) are mainly determined by the concentration of $^{235}U$, an almost linear relationship is shown between $^{235}U$ concentration and depletion time. The relationship, $c = 1$, for concentrations of $^{235}U$ and $^{238}U$ in Eq. (3) is originated from the homogeneous distribution at the BOL and verified in numerical results shown in Table 7. The analytical formula f(x,s) becomes thus:

$$f(x,s) = \begin{cases} b(x)s + 1, & ^{235}U \\ a(x)s^2 + b(x)s + 1, & ^{238}U \\ a(x)s^2 + b(x)s + c, & ^{239,241}Pu \end{cases} \quad (13)$$

The coefficients and corresponding fitting results of the above Equation are shown in Figure 11.

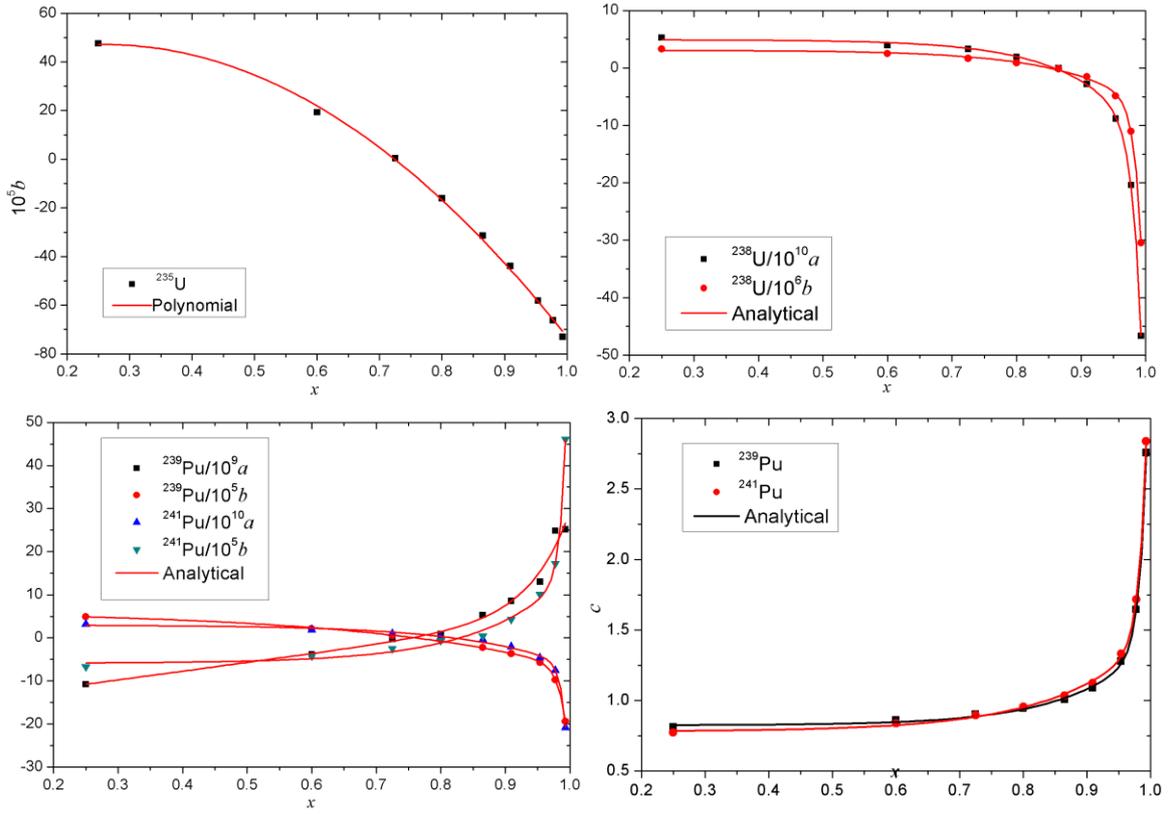

Figure 11. Second order polynomial fitting of $b$ for $^{235}$U concentrations, coefficients $a$ and $b$ for $^{238}$U concentrations, and coefficients $a$, $b$, and $c$ for $^{239,241}$Pu concentrations. The curves represent the fitting results, the scattering points are coefficients obtained in the fitting with fuel exposure.

$^{239}$Pu is the most important actinide besides $^{235}$U and $^{238}$U in the PWR. It mainly comes from $^{238}$U through the capture of one neutron and the two times of β decay. One of the main reasons for sharp power distribution close to fuel surface is the similar distribution of $^{239}$Pu. The neutron fission cross section of $^{239}$Pu is 1.9 times that of $^{235}$U at 0.1eV incident neutron energy according to JEFF-3.1.1 [31]. The fission reaction rate per nucleus of $^{239}$Pu is shown to be 2.5 times that of $^{235}$U in a PWR UO$_2$ fuel rod [20]. The previous study has also shown the similar radial distribution for $^{239}$Pu concentration and power [2]. Figure 12 shows the ratios of the relative concentrations of $^{239}$Pu to the present analytical formula. The values around 1.0 signify good description/prediction of the analytical formula. As shown in Figure 12, the five cases discussed in the present work have almost the same distribution of relative $^{239}$Pu concentration. There is about 5% deviation for the UO$_2$-zircaloy combination near the periphery. The $^{239}$Pu distribution is a little flatter in the reference case due to smaller neutron absorption cross section of zircaloy cladding. The UO$_2$/U$_3$Si$_2$-FeCrAl combination has flatter $^{239}$Pu distribution than U$_3$Si$_2$-FeCrAl combination but sharper distribution than UO$_2$-zircaloy because Si also has a contribution to hardening neutron spectrum. In fact, $^{16}$O is a double-magic nucleus, which leads to much smaller neutron absorption cross sections than most light elements in the fuel, such as Si and N. The substitution of other light elements in the UO$_2$ fuel causes the hardening of the spectrum and the sharper power distribution.

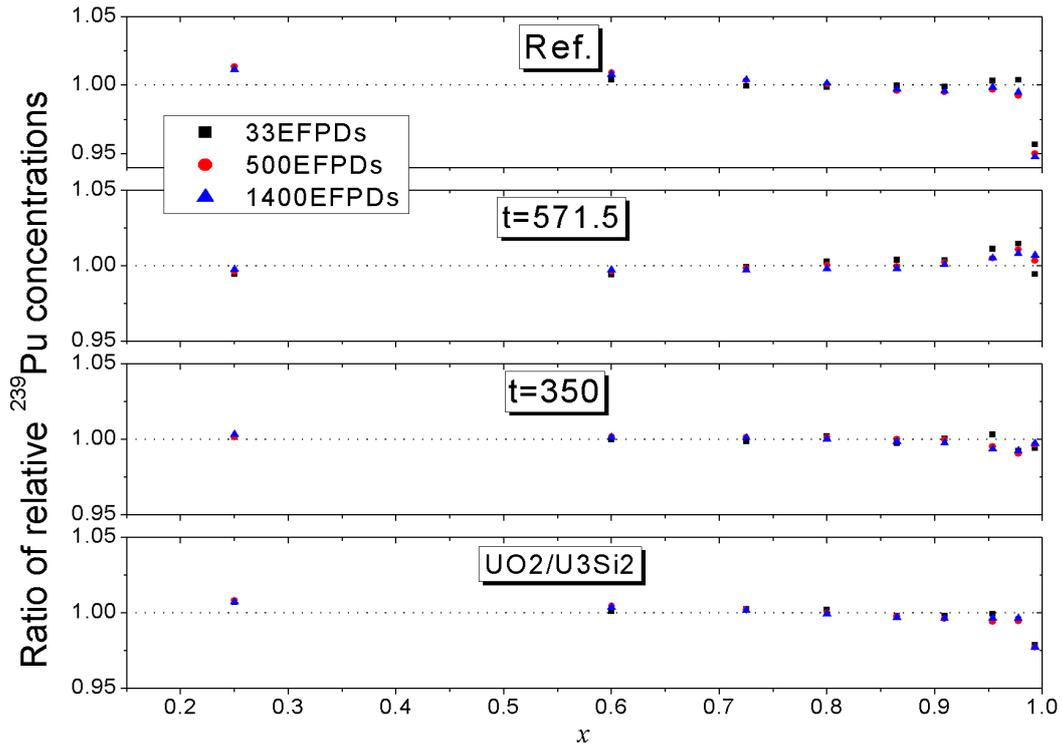

Figure 12. Ratios of relative $^{239}$Pu concentrations to those of t=450μm case

$^{241}$Pu is a product through a series of reactions in the fuel. The suggested analytical formula is not suitable to describe its concentration at the beginning of the operation but applicable after 100 EFPDs, as shown in Figure 10 (d). The relative $^{241}$Pu concentration distributions are quite similar for three cases of U$_3$Si$_2$-FeCrAl combinations. The deviations of $^{241}$Pu concentrations among the UO$_2$-zircaloy, U$_3$Si$_2$-FeCrAl, and UO$_2$/U$_3$Si$_2$-FeCrAl combinations are larger than those of $^{239}$Pu (ratio of the relative concentration of $^{241}$Pu ($^{239}$Pu) in UO$_2$-zircaloy to that in U$_3$Si$_2$-FeCrAl is 0.90 (0.95) at $x$=0.993). If Eq. (3) is used to calculate $^{239}$Pu and $^{241}$Pu concentration, it is better to perform a more careful fitting when high accuracy is required, such as the introduction of the fuel-cladding dependent term in the fitting.

3.5 Radial distributions of fuel temperature

Figure 13 shows the numerical results of the integral in the right-hand side of Eq. (10) for uniform power distribution case and the analytical radial power distributions at 300 EFPDs and 1400 EFPDs, respectively. The renormalization factor is 0.995 (0.989) to ensure the reliability of Eq. (11) at 300 EFPDs (1400 EFPDs).

If the convention is assumed to be the dominant heat transfer way between the cladding and moderator, the temperature at a certain position of the cladding and the moderator do not change in steady state with unchanged power. Therefore, the fuel temperature is lower in the case with realistic radial power distribution compared with uniform power distribution, of which the numerical results is shown in Figure 13. The periphery effect has a positive contribution to the decrement of the fuel temperature. Such effect may be more obvious in the annular fuel rod, such as the candidate ATF design proposed by Tan and Cai recently [32]. It demands further studies for both the cylindric and the annular fuel pellets.

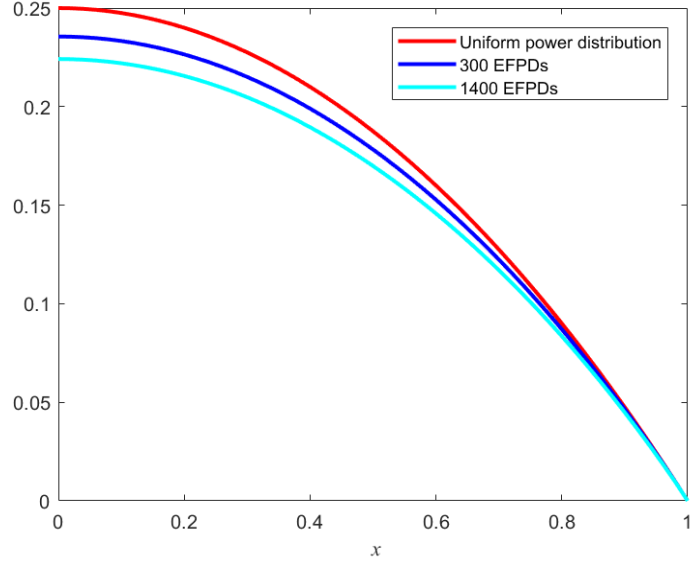

Figure 13. The right term in Eq. (10): $-\int_1^x \left[\int_0^{x'} x'' f(x'') dx''\right] / x' dx'$.

The right-hand side of the Eq. (10) is the same for 5 cases discussed in the present work, but the thermal conductibility of the fuel depends on material. Since the thermal conductivity of FeCrAl is lower than Zr-4, the temperatures at outer surface of fuels with FeCrAl cladding (the 4 candidate combinations) are higher than that with zircaloy cladding (reference case). In addition, the outer surface temperatures of fuels with 4 FeCrAl cladding cases increases with the cladding thickness because the gradient of temperature in the cladding is the same for the same heat transfer. Due to the much higher thermal conductivity, the temperature in the $U_3Si_2$ fuel is expected to be flatter (and lower in center) than that in $UO_2$ fuel. Quantitative results can be obtained with the empirical formulae (8) and (9), and the temperature at the surface of fuels. This conclusion is verified by multi-physics coupling analysis recently [33]. Figure 14 shows the numerical results in $UO_2$ and $U_3Si_2$-(571.5μm)FeCrAl respectively using 773 K and 830 K as the temperature at fuel outer surface, respectively. 773 K is a typical value for $UO_2$ fuel. 830 K for the $U_3Si_2$-(571.5μm)FeCrAl is used because Liu *et al.* [34] showed about 815 K and 825 K for 350μm and 450μm FeCrAl cladding. Higher temperature at fuel outer surface of $U_3Si_2$-FeCrAl combinations is in agreement with previous analyses. More accurate numerical results of $UO_2$ fuel can be obtained by considering the dependence of the thermal conductivity on other properties, including the burnup, the porosity, and the fission gas bubbles [35]. In addition, the temperature at the fuel outer surface depends on the fuel exposure [33], [34]. It is noticeable that the same radial power distribution for the 5 investigated cases leads to different fuel temperature distributions due to the different thermal conductivities of fuels and claddings and different thicknesses of claddings.

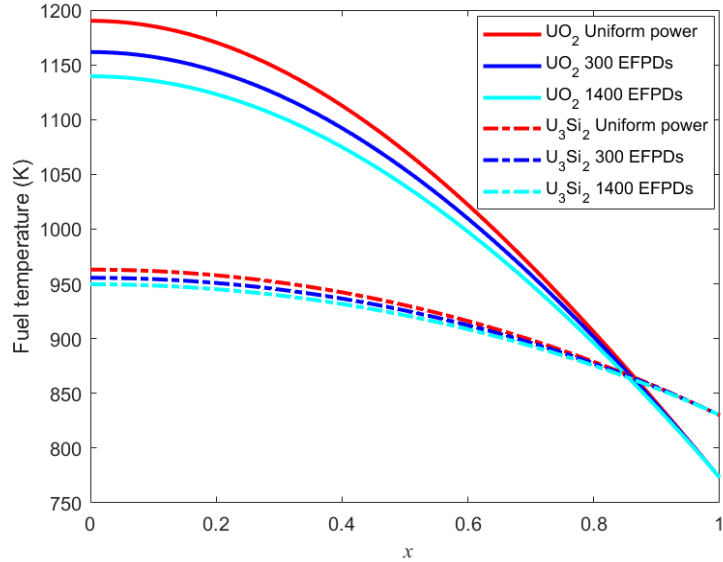

Figure 14. Fuel temperature distribution for $UO_2$ and $U_3Si_2$-(571.5µm)FeCrAl respectively using 773 K and 830 K as the temperature at fuel outer surface

## 4 Conclusions

If the same cycle length in a typical Westinghouse PWR is assumed, Monte Carlo simulations show very similar radial power distributions for the current $UO_2$-zircaloy system and four cases of candidate ATF with the FeCrAl cladding at any EFPD and radial distribution. Because of the different thermal conductivities of fuels and claddings and the different thicknesses of claddings, the present discussed five kinds of fuel-cladding combinations have different radial temperature distributions, although their radial power distributions are quite similar.

The analytical expression Eqs. (3) and (6) are proposed to describe the relative power distribution as a function of the fuel exposure and the radial position. The excellent fitting results on radial relative power distribution show that the analytical approximations can well describe the radial distributions of power and isotopic concentrations. The relative differences on the power distribution between Monte Carlo simulations and the suggested empirical formula are less than 3%. If the result from the outermost point in each case is neglected, the relative differences are within 1%. It should be noted that the outermost layer is rather thin, which introduces the largest simulation uncertainties among all layers with the same Monte Carlo setting. The relative burnup distributions can be calculated by integrating the obtained relative power distribution f($x,s$). For example, the relative deviations between the calculations from Monte Carlo and Eq. (5) are less than 2% at the EOL. It should be noted that the analytical expressions proposed in the present work are available at any possible depletion time between BOL and 2000 EFPDs. The present formulae are useful to reconstruct the isotopic concentrations at any position in a fuel pellet with the averaged concentrations from the neutronic calculation. They are also expected to provide a feedback to the macroscopic cross sections in neutron transport calculations due to the distributions of isotopic concentrations. The uncertainties of fitted coefficients can be used to investigate the systematic and the statistical uncertainties of the suggested function.

For example, the two kinds uncertainties of the simple nuclear mass formula are discussed through uncertainty decomposition method [36].

The numerical results show lower fuel temperature in realistic radial power distribution case. Compared with the homogeneous simplification of radial power distribution, the reactivity should be larger because of the Doppler effect. The Doppler effect describes the effect of the temperature on the width and height of resonance, which has an influence on neutronic parameters such as the resonance interface factor [37]. The present result provides a possible improvement for neutron transport code with analytical distribution on fuel temperature. The reactivity of the $U_3Si_2$-FeCrAl case is expected to be higher in a real situation due to the overestimation of fuel temperature in previous neutronic calculations, including ours [2]. The flatter (and lower in fuel center) fuel temperature in $U_3Si_2$ fuel, compared with the normal case in the $UO_2$ fuel, is shown in a recent multi-physics coupling study [33]. In addition, the fuel temperature increases with the cladding thickness for the same fuel type and the same cladding materials. The present suggested analytical formula should be useful in the further neutron transport and multi-physics coupling calculations.

The present work provides useful information for further fuel performance studies, such as the fuel expansion, the thermal creep in the fuel, and the fission product swelling, because these properties depend on the fuel temperature. These effects induced by the variation of fuel temperature have also influences on mechanical properties.

## Acknowledgments

The authors acknowledge the authorized usage of the RMC code from Tsinghua University for this study. This work has been supported by the National Natural Science Foundation of China under Grant No. 11775316, the Tip-top Scientific and Technical Innovative Youth Talents of Guangdong special support program under Grant No. 2016TQ03N575, and the Fundamental Research Funds for the Central Universities under Grant No. 17lgzd34.

## Appendix

Coefficients and associated uncertainties in Eq. (3) for radial power and isotopic concentration distribution, and coefficients in Eq. (6) for power distribution are given in following.

Table 5 Coefficients and associated uncertainties obtained by fitting and the coefficient of determination for fitting at each position for relative power distribution

| $x$ | $10^9 a(x)$ | | $10^5 b(x)$ | | $c(x)$ | | $R^2$ |
|---|---|---|---|---|---|---|---|
| 0.2500 | 7.47 | 19.4% | -5.18 | 5.19% | 0.94 | 0.08% | 0.994 |
| 0.6000 | 9.79 | 9.30% | -5.63 | 2.98% | 0.97 | 0.05% | 0.998 |
| 0.7250 | 11.88 | 7.00% | -5.71 | 2.69% | 1.00 | 0.04% | 0.998 |
| 0.8000 | 12.07 | 4.50% | -5.07 | 1.98% | 1.02 | 0.03% | 0.999 |
| 0.8650 | 10.12 | 7.20% | -3.52 | 3.86% | 1.03 | 0.04% | 0.993 |
| 0.9090 | 5.46 | 13.0% | -0.44 | 29.8% | 1.05 | 0.04% | 0.943 |

| 0.9536 | -10.96 | 17.4% | 8.77 | 4.03% | 1.08 | 0.09% | 0.997 |
| 0.9775 | -51.54 | 8.70% | 29.12 | 2.87% | 1.10 | 0.21% | 0.998 |
| 0.9930 | -205.6 | 6.10% | 93.54 | 2.49% | 1.12 | 0.58% | 0.998 |

Table 6 Coefficients of Eq. (6) for radial power distribution

|  | A | B | C | D | E | $R^2$ |
|---|---|---|---|---|---|---|
| $10^9 a(x)$ | -372.6 | -146.7 | -105.3 | -34.74 | 10.40 | 0.9993 |
| $10^5 b(x)$ | 37.00 | -21.79 | 146.7 | -111.3 | -5.530 | 1.0000 |
| $c(x)$ | 0.1670 | -2.565 | 0.04993 | -25.97 | 0.9147 | 0.9943 |

Table 7 Coefficients and associated uncertainties obtained by fitting at each position for relative isotopic concentration distribution for $^{235,238}$U, and $^{239,241}$Pu

|  | $x$ | $10^9 a(x)$ |  | $10^5 b(x)$ |  | $c(x)$ |  |
|---|---|---|---|---|---|---|---|
| $^{235}$U | 0.2500 | 0 | 0 | 476.0 | 0.52% | 1.000 | 0.02% |
|  | 0.6000 | 0 | 0 | 192.9 | 0.27% | 1.000 | 0.00% |
|  | 0.7250 | 0 | 0 | 03.35 | 26.7% | 1.000 | 0.01% |
|  | 0.8000 | 0 | 0 | -160.4 | 1.05% | 1.000 | 0.01% |
|  | 0.8650 | 0 | 0 | -314.3 | 0.38% | 1.000 | 0.01% |
|  | 0.9090 | 0 | 0 | -439.0 | 0.14% | 1.000 | 0.01% |
|  | 0.9536 | 0 | 0 | -581.4 | 0.12% | 1.000 | 0.01% |
|  | 0.9775 | 0 | 0 | -661.6 | 0.17% | 1.000 | 0.01% |
|  | 0.9930 | 0 | 0 | -730.8 | 0.29% | 1.000 | 0.02% |
| $^{238}$U | 0.2500 | 52.9 | 3.62% | 3.28 | 1.10% | 1.000 | 0.00% |
|  | 0.6000 | 39.4 | 5.02% | 2.50 | 1.49% | 1.000 | 0.00% |
|  | 0.7250 | 33.3 | 8.66% | 1.67 | 3.25% | 1.000 | 0.00% |
|  | 0.8000 | 18.9 | 14.0% | 0.90 | 5.51% | 1.000 | 0.00% |
|  | 0.8650 | -0.50 | 514% | -0.14 | 38.6% | 1.000 | 0.00% |
|  | 0.9090 | -27.8 | 10.1% | -1.54 | 3.41% | 1.000 | 0.00% |
|  | 0.9536 | -88.2 | 4.75% | -4.86 | 1.62% | 1.000 | 0.00% |
|  | 0.9775 | -203.8 | 1.52% | -11.05 | 0.53% | 1.000 | 0.00% |
|  | 0.9930 | -466.5 | 0.73% | -30.43 | 0.21% | 1.000 | 0.00% |
| $^{239}$Pu | 0.2500 | -10.76 | 3.33% | 4.94 | 1.42% | 0.811 | 0.03% |
|  | 0.6000 | -3.80 | 15.0% | 2.14 | 5.22% | 0.860 | 0.04% |
|  | 0.7250 | -0.30 | 86.5% | 0.51 | 10.1% | 0.901 | 0.02% |
|  | 0.8000 | 0.88 | 66.3% | -0.45 | 25.2% | 0.944 | 0.04% |
|  | 0.8650 | 5.29 | 10.6% | -2.25 | 4.90% | 1.006 | 0.04% |
|  | 0.9090 | 8.58 | 6.20% | -3.67 | 2.84% | 1.087 | 0.03% |
|  | 0.9536 | 13.0 | 6.90% | -5.74 | 3.08% | 1.278 | 0.05% |
|  | 0.9775 | 24.88 | 4.20% | -9.72 | 2.12% | 1.644 | 0.04% |
|  | 0.9930 | 25.15 | 3.78% | -19.42 | 0.96% | 2.756 | 0.02% |
| $^{241}$Pu | 0.2500 | 31.96 | 4.91% | -6.68 | 4.94% | 0.773 | 0.16% |

|   | 0.6000 | 18.87  | 5.81% | -4.19 | 5.50% | 0.838 | 0.10% |
|   | 0.7250 | 10.30  | 30.2% | -2.49 | 26.2% | 0.895 | 0.28% |
|   | 0.8000 | 1.36   | 106%  | -0.59 | 51.0% | 0.955 | 0.12% |
|   | 0.8650 | -4.27  | 47.0% | 0.45  | 94.6% | 1.036 | 0.16% |
|   | 0.9090 | -20.55 | 6.20% | 4.30  | 6.26% | 1.125 | 0.09% |
|   | 0.9536 | -45.96 | 5.98% | 10.10 | 5.72% | 1.333 | 0.17% |
|   | 0.9775 | -75.2  | 4.94% | 17.28 | 4.52% | 1.718 | 0.17% |
|   | 0.9930 | -208.0 | 1.79% | 46.09 | 1.70% | 2.838 | 0.10% |